\documentclass[aps,prl,twocolumn,superscriptaddress]{revtex4}
\usepackage{bm}
\usepackage{graphicx,amsmath,latexsym,amssymb}
\usepackage{color}
\usepackage{dcolumn}

\newcommand{\bra}[1]{\left\langle\!#1 \right|}
\newcommand{\ket}[1]{\left|#1\!\right\rangle}

\begin{document}

\title{{Quantum Electrodynamics of Casimir Momentum: \\
Momentum of the Quantum Vacuum?}}

\author{S. Kawka}
\affiliation{Laboratoire de Physique et Mod\'{e}lisation des Milieux
Condens\'{e}s, Universit\'{e} Joseph Fourier and CNRS, Maison des
Magist\`{e}res, BP 166, 38042 Grenoble, France}

\author{B.A. van Tiggelen}

\affiliation{Laboratoire de Physique et Mod\'{e}lisation des Milieux
Condens\'{e}s, Universit\'{e} Joseph Fourier and CNRS, Maison des
Magist\`{e}res, BP 166, 38042 Grenoble, France}

\date{November 3, 2009}

%\pacs{12.20.-m}{Quantum electrodynamics}
%\pacs{03.70.+k}{Theory of quantized fields}
%\pacs{11.10.Gh}{Renormalization}

\begin{abstract}
The electromagnetic vacuum is known to have energy. It has been recently argued that the quantum vacuum can possess momentum, that adds up to the momentum of matter. This ``Casimir momentum'' is closely related to the Casimir effect, in which case energy is exchanged. In previous theory it was treated semi-classically. We present a non-relativistic quantum theory for the linear momentum of electromagnetic zero-point fluctuations, considering an harmonic oscillator subject to crossed, quasi-static magnetic and electric and coupled to the quantum vacuum. We derive a contribution of the quantum vacuum to the linear pseudo-momentum and give a new estimate for the achievable speed. Our analysis show that the effect exists and that it is finite.
\end{abstract}

\maketitle

Casimir energy refers to the electromagnetic (EM) energy that shows
up when dielectric or metallic objects interact with the quantum
vacuum. It is undoubtedly one of the most fascinating phenomena in
physics, with a rich history in the 20Th century. Casimir forces
become important on sub-micron scales and are thus believed to play
an important role in nano-optics \cite{lambrecht}. Casimir energy
has been the subject of many speculations, such as its role in
sonoluminescence \cite{sono1} or in the cosmological constant
problem \cite{cosmo}.

The standard Casimir effect refers to the reduction of the EM
zero-point energy when two ideal metallic plates approach \cite{casimir}. Other
well-known phenomena related to Casimir energy are Van Der Waals and
Casimir-Polder forces between neutral atoms \cite{polder}, the Lifshitz forces
between dielectric media, and arguably the most famous among all,
the Lamb shift of atomic levels. Shortly after its observation by
Lamb in 1947 \cite{lamb}, Bethe  explained the Lamb shift by the change in EM
vacuum energy caused by the interaction of the atom with the quantum
vacuum \cite{bethe,milonni}. The Lamb shifts in light atoms are now understood to be
basically nonrelativistic QED phenomena, although full relativistic
theory, including the contribution of several percents due to vacuum
polarization, is necessary to come to the extraordinary agreement
with experiment, unprecedented in physics. For the two-photon 1S-2S
transition in atomic hydrogen, the shift is known up to several
cycles \cite{hansch}.

Energy and momentum are naturally related by relativity. The search
for ``Casimir momentum'' seems therefore obvious. In 2004 Feigel \cite{feigel}
proposed a quantum correction to the momentum of dielectric media
exposed to static electric and magnetic fields. In this case, classical
electrodynamics provides the following expression for the
linear momentum of a neutral, polarizable object with mass $M$,
{
\begin{equation}
\mathbf{Q}_{class} = M\mathbf{v} - \alpha(0) \mathbf{E}_0 \times
\mathbf{B}_0 \label{class}
\end{equation}}
which is conserved in time, even if the external electric field
$\mathbf{E}_0$ is varied slowly in time. Here $\alpha(0)$ is the
static polarizability, with the dimension of a volume. The
semi-classical theory of Ref.\cite{feigel} predicts a strongly diverging contribution of
the quantum vacuum to Eq.~(\ref{class}), quite similar to the one
encountered for Casimir energy. Fortunately, spatial gradients of
Casimir energy - observable as forces - are often found not to
diverge. Momentum however is an observable parameter and the
divergence does pose a problem. It has been suggested that UV
divergences are not physical and should disappear into the  values
attributed to physical observables, such as inertial mass, electric
charge or cosmological constant \cite{milton}. If this is true it is
not evident that the prediction of ``Casimir momentum'' found in Ref.\cite{feigel} will survive or be measurable. An obvious next question is
what physical observable will then absorb the UV divergence of
Casimir momentum. In this work we provide first answers to these
questions. We use the method of mass renormalization first employed
by Bethe and Kramers  that results in a finite Casimir momentum of
simple quantum objects.

 It is now realized that Casimir momentum emerges quite
generally in so-called bi-anisotropic media, in which also magnetic
fields can induce an electric polarization {\cite{brevik,bart}}. Except in media exposed
to external EM fields, bi-anisotropy also occurs in moving
dielectric media. This follows from the relativistic transformations
of EM fields, and shall be discussed elsewhere \cite{kawka2}. In
general, like spin, bi-anisotropic behavior and Casimir momentum can
be viewed as ``remnants'' of special relativity \cite{comment} in
non-relativistic theory that often suffices to describe phenomena
quantitatively.

{We consider here the following system: a 3D harmonic oscillator - composed of two particles with opposite charge $q_1= + e$ and $q_2= -e$ and masses
$m_i$ - exposed to crossed, homogeneous static EM fields $\mathbf{E}_0, \mathbf{B}_0$ which constitute our bi-anisotropic object, coupled to the EM vacuum. In the Schr\"{o}dinger picture the Hamiltonian is given by
\begin{equation}\label{htot}
\begin{split}
H &= \sum_{i=1}^2 \left[ \frac{1}{2m_i}\big(\mathbf{p}_i - q_i \mathbf{A}_t(\mathbf{r}_i)\big)^2 - q_i \mathbf{E}_0.\mathbf{r}_i \right] +  V(\mathbf{r}) \\
&+ \sum_{\mathbf{k}\boldsymbol{\epsilon}} \hbar \omega_{k}       \left[ a_{\mathbf{k} \boldsymbol{\epsilon}}^\dag a_{\mathbf{k}\boldsymbol{\epsilon}} + \frac{1}{2} \right]
\end{split}
\end{equation}
where $\mathbf{A}_t= \mathbf{A}_0 +\mathbf{A}$ is the total vector potential containing a contribution from the external, static, classical magnetic field, described by the classical vector potential
$\mathbf{A}_0(\mathbf{r})= \frac{1}{2} \mathbf{B}_0 \times
\mathbf{r}$, and the quantum operator $\mathbf{A}$ of the EM field. The EM bath will be treated in the Coulomb gauge.
We will use $\mathbf{R}=(m_1 \mathbf{r}_1+ m_2\mathbf{r}_2)/M$ and
$\mathbf{r}=\mathbf{r}_1-\mathbf{r}_2$ for the center of mass
position and the inter-particle distance, with \emph{conjugate}
momenta $\mathbf{P}$ and $\mathbf{p}$, respectively; $M=m_1+m_2$ and $\mu=(1/m_1+1/m_2)^{-1}$ are the total and reduced mass. For a harmonic oscillator we can then write $V(\mathbf{r})=\frac{1}{2}\mu \omega_0^2 \mathbf{r}^2$.

As we are looking for momentum, we notice the existence of a conserved \emph{pseudo}-momentum $\mathbf{K}$ that commutes with $H$ \cite{cohen}, even when the electric field is varied in time. The electric field is a parameter that can be varied experimentally. This momentum has contributions from both atom and radiation,
\begin{equation}\label{Kcanoniq}
    \mathbf{K}= \mathbf{P}  + \frac{e}{2}\mathbf{B}_0 \times \mathbf{r} + \sum_{\mathbf{k}\boldsymbol{\epsilon}} \hbar
    \mathbf{k}\left[ a_{\mathbf{k} \boldsymbol{\epsilon}}^\dag
    a_{\mathbf{k}\boldsymbol{\epsilon}} + \frac{1}{2} \right]
\end{equation}
It is a \emph{pseudo}-momentum as it is a constant of the motion only if the external magnetic field $\mathbf{B}_0$ is time independent and homogeneous, as we will assume here. Note that we are interested in the change of the total \emph{kinetic} momentum $\mathbf{P}_{\mathrm{kin}}$ of the oscillator in the presence of vacuum, which is not equal to $\mathbf{K}$, neither to $\mathbf{P}$. Yet, because $\mathbf{K}$ is conserved, even for a slowly time-dependent electric field $\mathbf{E}_0$, it is the appropriate momentum to look at. In the presence of magnetic fields the \emph{kinetic} and \emph{conjugate} momenta operators are related by $\mathbf{P}_{\mathrm{kin}}= \mathbf{P} - e \Delta \mathbf{A}_t$ with $e\Delta \mathbf{A}_t= e\mathbf{A}_t(\mathbf{r}_1)-e\mathbf{A}_t(\mathbf{r}_2)$, so that we can obtain 
\begin{equation}\label{K}
    \mathbf{K}= \mathbf{P}_{\mathrm{kin}}  + e\mathbf{B}_0 \times \mathbf{r} + e \Delta \mathbf{A} 
    +\sum_{\mathbf{k}\boldsymbol{\epsilon}} \hbar
    \mathbf{k}\left[ a_{\mathbf{k} \boldsymbol{\epsilon}}^\dag
    a_{\mathbf{k}\boldsymbol{\epsilon}} + \frac{1}{2} \right]
\end{equation}
The operator $e\Delta \mathbf{A}= e\mathbf{A}(\mathbf{r}_1)-e\mathbf{A}(\mathbf{r}_2)$ guarantees the gauge-invariant
contribution of the ``longitudinal'' vacuum field to the
pseudo-momentum,  in terms of the  vector potential
$\mathbf{A}(\mathbf{r})$ quantized as usual inside a quantization
volume $V$, $\mathbf{A}(\mathbf{r})=\sum_{\mathbf{k}\boldsymbol{\epsilon}}    \mathcal{A}_\mathbf{k}  \boldsymbol{\epsilon}  \left[a_{\mathbf{k}\boldsymbol{\epsilon}}e^{i\mathbf{k}\mathbf{r}} + a_{\mathbf{k} \boldsymbol{\epsilon}}^\dag e^{-i\mathbf{k}\mathbf{r}} \right]$. We will be obliged to go beyond the electric dipole approximation -in which $e\Delta \mathbf{A}$ would be neglected- to treat the high wave numbers of zero-point
fluctuations accurately. The last term in Eq.~(\ref{K}) stems from the
 ``transverse'' electromagnetic field in the vicinity of the atom \cite{cohen}.

We wish to express the expectation value $\overline{\mathbf{K}}= \langle \Psi_0 |
\mathbf{K} | \Psi_0 \rangle$, of the pseudo-momentum in the total ground state
$| \Psi_0 \rangle $ in terms of the two relevant vectors: the kinetic momentum $M \mathbf{v}$ of the oscillator and the
magneto-electric vector $\mathbf{E}_0 \times \mathbf{B}_0$. We will use perturbation theory in the coupling between the magnetoelectric oscillator and the EM field, where the small parameter of the expansion is the fine structure constant $\alpha$. To lowest order, only the emission and subsequent re-absorption of one virtual photon will contribute at this one-loop level of the theory, which will thus be second order perturbation theory. In order to facilitate perturbation theory, we split the Hamiltonian (\ref{htot}) up as follow,
\begin{gather}\label{hdippel}
H = H_0 + H_F + W \\
\nonumber H_0 = \sum_{i=1}^2 \left[ \frac{1}{2m_i}\big(\mathbf{p}_i - q_i \mathbf{A}_0(\mathbf{r}_i)\big)^2 - q_i \mathbf{E}_0.\mathbf{r}_i \right] +  \frac{1}{2}\mu \omega_0^2 \mathbf{r}^2\\
\nonumber H_F = \sum_{\mathbf{k}\boldsymbol{\epsilon}} \hbar \omega_{k}       \left[ a_{\mathbf{k} \boldsymbol{\epsilon}}^\dag a_{\mathbf{k}\boldsymbol{\epsilon}} + \frac{1}{2} \right]\\
\nonumber W = \sum_{i=1}^2 -\frac{q_i}{m_i}(\mathbf{p}_i-q_i\mathbf{A}_{0}(\mathbf{r}_i))\mathbf{A}(\mathbf{r}_i) + \frac{q_i^2}{2m_i}\mathbf{A}(\mathbf{r}_i)^2
\end{gather}
The operator $W$ represents the perturbation of the quantum vacuum on the atom. The term $\mathbf{A}^2$ can be disregarded because it does not couple field and matter, and at this order its contribution to $\overline{\mathbf{K}}$ will vanish due to basic selection rules. The photon field in free space described by $H_F$ is well known.} Finally, the quantum-mechanics of a 3D harmonic oscillator exposed to crossed, homogeneous static EM fields, described by $H_0$, was discussed in detail and non-perturbationally by Dippel etal \cite{dippel}.
{$H_0$ and $H_F$ act in different Hilbert-spaces and the basis will be the direct product of their eigenvectors. As we are interested in the center-of-mass motion of the 3D harmonic oscillator, one key point is the distinction between the center-of-mass and the internal coordinates. Without the quantum vacuum, the pseudo-momentum $\mathbf{K}$ reduces to,
\begin{equation}\label{Q}
    \mathbf{Q} = \mathbf{P}_{\mathrm{kin}} + e\mathbf{B}_0 \times \mathbf{r}
\end{equation}
which commutes with the atomic Hamiltonian. $\mathbf{Q}$ is the quantum-mechanical operator corresponding to the classical pseudo-momentum $\mathbf{Q}_{class}$ in Eq.~(\ref{class})}. For the purpose of this work it is convenient to choose eigenfunctions that simultaneously diagonalize the
atomic Hamiltonian and the pseudo-momentum $\mathbf{Q}$, labeled by
the eigenvalue $\mathbf{Q}_0$, {as $\mathbf{Q}$ appears directly in $\mathbf{K}$ as seen by inserting Eq.~(\ref{K}) into Eq.~(\ref{Q})}. It will be sufficient
to ignore all contributions other than on those linear in either
$\mathbf{E}_0$ and $\mathbf{B}_0$, as indicated by the sign
$\asymp$. In this approximation, the magneto-electric oscillator is unitary equivalent
to an  isotropic harmonic oscillator as expressed by,
\begin{widetext}
\begin{equation}\label{ho}
|\mathbf{n}, \mathbf{Q}_0\rangle  \asymp \exp\left(\frac{i}{\hbar}\mathbf{Q}_0\cdot \mathbf{R}\right)
 \exp\left[-\frac{i}{2\hbar}(\mathbf{B}_0 \times \mathbf{r})\cdot \mathbf{R}\right]
 \exp\left(-\frac{i}{\hbar}\mathbf{p}_0\cdot \mathbf{r}\right) \exp\left(-\frac{i}{\hbar}\mathbf{p}\cdot
 \mathbf{r}_0\right)
|\phi_\mathbf{n} \rangle
\end{equation}
\end{widetext}

\noindent $\mathbf{n} = (n_x, n_y, n_z)$, $n_i=0, 1, \cdots$ denotes
the quantum levels of the oscillator. The first two exponentials on
the right denote translational momentum of the center of mass, and
governed by the conjugate momentum $\mathbf{P}$. The last pair of
exponentials eliminate the static electric field from the picture,
with the eigenfunctions of the oscillator shifted out of the center
of mass over a distance $\mathbf{r}_0= e^{-1}\alpha(0)
(\mathbf{E}_0+ \mathbf{Q}_0 \times\mathbf{ B}_0/M)$, and the reduced
momentum shifted by
$\mathbf{p}_0\asymp(2M)^{-1}(m_2-m_1)\alpha(0)(\mathbf{E}_0\times
\mathbf{B}_0)$ with $\alpha(0) = e^2/\mu \omega_0^2$ the static
polarizability of the oscillator. {This last feature is reminiscent of bi-anisotropic activity, and will generate a dominant contribution to $\overline{\mathbf{K}}$. Note that it vanishes for $m_1=m_2$.} Due to the static magnetic field,
the oscillator states $|\phi_\mathbf{n} \rangle$ are in principle still
anisotropic and even in $B_0$.  The anisotropy is estimated by the
small parameter $eB_0 a/ (\hbar/a) \approx 10^{-5}$, with $a$
the atomic size. This anisotropy constitutes corrections nonlinear in the
applied fields to the final result for the total momentum. We can
therefore neglect it.

Upon taking the quantum-expectation value of Eq.~(\ref{Q}) for the
atomic ground state, {which still ignores the quantum vacuum,} reveals that the eigenvalue $\mathbf{Q}_0$  is
just equal to the classical expression~(\ref{class}). The total
energy of the oscillator in the ground state $E_0 \asymp
\frac{3}{2}\hbar \omega_0 + Q_0^2/2M$ is minimal when
$\mathbf{Q}_0=0$, i.e. for a \emph{finite} kinetic momentum.

In the absence of the interaction with the quantum vacuum, the eigenstates
are just the direct products $|\mathbf{n}, \mathbf{Q}_0, \mathbf{n}_\mathbf{k} \rangle =
|\mathbf{n}, \mathbf{Q}_0\rangle \otimes | \mathbf{n}_\mathbf{k} \rangle$, with
unperturbed energies $E_{\mathbf{nQ_0}\mathbf{n_k}} =
E_{\mathbf{nQ}_0} + \sum_\mathbf{k} \hbar \omega_\mathbf{k}
(n_\mathbf{k}+ \frac{1}{2})$. Here $\mathbf{n}_\mathbf{k} $ is
the occupation of the EM Fock states with photon momentum $\hbar
\mathbf{k}$. The ground state $| \Psi_0 \rangle $ follows from second-order perturbation in the
coupling $W$ to the quantum vacuum ($\sum^{'}$ avoids zeros in the denominator),
\begin{eqnarray}
\label{groundstate}
\nonumber \ket{\Psi_{0}} = \ket{\mathbf{0},\mathbf{Q}_0,\mathbf{0}} + \sum_{\mathbf{l}\mathbf{Q}\mathbf{n}} { }^{'} \frac{ W_{\mathbf{lQn},\mathbf{0}\mathbf{Q}_0\mathbf{0}} }{E_{\mathbf{0}\mathbf{Q}_0\mathbf{0}}-E_{\mathbf{lQn}}}  \ket{\mathbf{l},\mathbf{Q},\mathbf{n}}
\\
\nonumber + \sum_{\mathbf{lQn}}  { }^{'}\sum_{\mathbf{s}\mathbf{Q}'\mathbf{m}}  { }^{'} \frac{ W_{\mathbf{lQn},\mathbf{s}\mathbf{Q}'\mathbf{m}} W_{\mathbf{s}\mathbf{Q}'\mathbf{m},\mathbf{0}\mathbf{Q}_0\mathbf{0}}}{(E_{\mathbf{0}\mathbf{Q}_0\mathbf{0}}-E_{\mathbf{lQn}})(E_{\mathbf{0}\mathbf{Q}_0\mathbf{0}}-E_{\mathbf{s}\mathbf{Q}'\mathbf{m}})} \ket{\mathbf{l},\mathbf{Q},\mathbf{n}}
\\
\nonumber + \sum_{\mathbf{lQn}}  { }^{'}\frac{| W_{\mathbf{lQn},\mathbf{0}\mathbf{Q}_0\mathbf{0}} |^2}{(E_{\mathbf{0}\mathbf{Q}_0\mathbf{0}}-E_{\mathbf{lQn}})^2}  \ket{\mathbf{l},\mathbf{Q},\mathbf{n}}
\end{eqnarray}
Only the emission and subsequent re-absorption of one virtual photon is considered, which generates a temporary recoil momentum $\mathbf{Q}_0-\hbar \mathbf{k}$ of the oscillator,
\begin{align}
\label{Kperturb}
&\overline{\mathbf{K}}  = \mathbf{Q}_0  \\
&\nonumber  + \! e^22\mathrm{Re }\sum_{\mathbf{lk}\boldsymbol{\epsilon}} \mathcal{A}_{\mathbf{k}}^{2}\boldsymbol{\epsilon} \, \frac{ \bra{\phi_{\mathbf{0}}} (e^{i\mathbf{k}(\mathbf{r}+\mathbf{r}_0)\frac{m_2}{M}} - \! e^{-i\mathbf{k}(\mathbf{r}+\mathbf{r}_0)\frac{m_1}{M}}) \ket{\phi_{\mathbf{l}}}  \Omega_{\mathbf{l},\mathbf{0}}^*  }{ E_{\mathbf{0}\mathbf{Q}_0\mathbf{0}}- \! E_{\mathbf{l}(\mathbf{Q}_0-\hbar \mathbf{k})1_{k}}} \\
& \nonumber  +  \!  \mathbf{B}_0 \times \!  2\mathrm{Re }\sum_{\mathbf{lsk}\boldsymbol{\epsilon}} \frac{ e^3\mathcal{A}^{2}_{\mathbf{k}}\bra{\phi_{\mathbf{0}}} \mathbf{r}+\mathbf{r}_0 \ket{\phi_{\mathbf{l}}} \Omega_{\mathbf{l},\mathbf{s}} \Omega_{\mathbf{s},\mathbf{0}}^*  }{(E_{\mathbf{0}\mathbf{Q}_0\mathbf{0}}-E_{\mathbf{l}\mathbf{Q}_0\mathbf{0}})( E_{\mathbf{0}\mathbf{Q}_0\mathbf{0}}-E_{\mathbf{s}(\mathbf{Q}_0-\hbar \mathbf{k})1_{k}})} \\ 
&\nonumber  +  \! \mathbf{B}_0 \times  \! \sum_{\mathbf{lsk}\boldsymbol{\epsilon}} \frac{ e^3\mathcal{A}^{2}_{\mathbf{k}}\bra{\phi_{\mathbf{s}}} \mathbf{r}+\mathbf{r}_0 \ket{\phi_{\mathbf{l}}}  \Omega_{\mathbf{l},\mathbf{0}}^*\Omega_{\mathbf{0},\mathbf{s}} }{(E_{\mathbf{0}\mathbf{Q}_0\mathbf{0}}- \! E_{\mathbf{l}(\mathbf{Q}_0-\hbar \mathbf{k})1_{k}})( E_{\mathbf{0}\mathbf{Q}_0\mathbf{0}}- \! E_{\mathbf{s}(\mathbf{Q}_0-\hbar \mathbf{k})1_{k}} )} 
\end{align}
{The second term on the right-hand-side is due to $e\Delta \mathbf{A}$ in Eq.~(\ref{K}), the two last stem from $e\mathbf{B}_0\times\mathbf{r}$. We introduced the matrix element $\Omega_{\mathbf{l},\mathbf{s}} = \bra{\phi_{\mathbf{l}}} \Omega \ket{\phi_{\mathbf{s}}}$ of the operator}
\begin{eqnarray}\label{Omega}
\nonumber \Omega &=&  e\boldsymbol{\epsilon} \cdot [\mathbf{B}_0 \times (\mathbf{r}+\mathbf{r}_0) ] \left(\frac{e^{i\mathbf{k}(\mathbf{r}+\mathbf{r}_0)\frac{m_2}{M}}}{m_1} -\frac{e^{-i\mathbf{k}(\mathbf{r}+\mathbf{r}_0)\frac{m_1}{M}}}{m_2}\right)  \\
\nonumber &-& \frac{\mathbf{Q}_0}{M} \cdot  \boldsymbol{\epsilon} \left(e^{i\mathbf{k}(\mathbf{r}+\mathbf{r}_0)\frac{m_2}{M}} -e^{-i\mathbf{k}(\mathbf{r}+\mathbf{r}_0)\frac{m_1}{M}}\right)  \\
\nonumber  &-&(\mathbf{p}-\mathbf{p}_0) \cdot \boldsymbol{\epsilon} \left(\frac{e^{i\mathbf{k}(\mathbf{r}+\mathbf{r}_0)\frac{m_2}{M}}}{m_1} +\frac{e^{-i\mathbf{k}(\mathbf{r}+\mathbf{r}_0)\frac{m_1}{M}}}{m_2}\right)
\end{eqnarray}
and $\mathcal{A}^2_\mathbf{k} = \hbar/2\varepsilon_0 V kc$ familiar from quantum optics. {$\Omega$ stems directly from the development of $W$ applied to a one photon transition.}

Three kinds of contributions to
$\overline{\mathbf{K}}$ can be identified. The first class is proportional to
$\mathbf{Q}_{0}$ (see the middle term in $\Omega_{\mathbf{l},\mathbf{s}}$) that survives even
in the absence of external fields. {This term can be seen to affect the inertial mass of the atom, by typically $\frac{3}{2}\frac{\hbar \omega_0}{c^2}$ in accordance with the equivalence principle of energy and inertia} \cite{kawka2}. Its UV divergence can
be absorbed into the total mass $M$ in the same way as will be discussed below for the ME divergences.

The second class, represented by the two last contributions in
Eq.~(\ref{Kperturb}), are actually QED contributions to the induced
electrical dipole moment $\langle \Psi_0 | e\mathbf{r} | \Psi_0
\rangle$ of the oscillator, that find their way to the total
momentum via the classical expression~(\ref{class}). It is
straightforward to calculate these corrections - they actually do
not diverge and they are relatively small - but we note that if an
experimental value for $\alpha(0)$ is used to evaluate the
``classical'' contribution, these terms are automatically included. In
this sense they do not constitute a genuine ``Casimir momentum''.

The term $e \Delta \mathbf{A}$ in Eq.~(\ref{K}) is a genuine contribution of the vacuum radiation to the
pseudo-momentum. It will be seen
 to generate a momentum linear in $\mathbf{E}_0 \times \mathbf{B}_0$ by means of the second term in Eq.~(\ref{Kperturb}) and the third term in $\Omega$. The following calculation will focus
 on this contribution. It suffers from a UV
 divergence, that can be eliminated  by exactly the same mass
 regularization as
 applied by Bethe in his calculation of the Lamb shift \cite{bethe}. In
 particular, this procedure establishes that the reduced mass featuring in the static
 polarizability $\alpha(0) = e^2/\mu \omega_0^2$ will be replaced by the
 ``observed'' reduced mass. It is straightforward to show that the
  so-called transverse
 electromagnetic momentum, represented by the last term of
Eq.~(\ref{K}), does not contribute a net momentum. This follows from selection rules and spatial symmetry.

We apply the closure relation $\sum_\mathbf{l}
\ket{\phi_{\mathbf{l}}}\bra{\phi_{\mathbf{l}}}  ({E(k) +
E_\mathbf{l}})^{-1} =  ({E(k) + H_{ho}})^{-1}$ in the second
contribution to $\overline{\mathbf{K}}$. {This generates two terms
involving exponentials with opposite phases that cancel. The contribution from the first term in $\Omega$ to $\overline{\mathbf{K}}$ is 
\begin{equation}\label{dvp1}
 \alpha(0) \mathbf{E}_0 \times \mathbf{B}_0   \, \,
\frac{4 \alpha}{3\pi}  \left[ \frac{\hbar^2}{m_i} \int \frac{kdk}{
\frac{ \hbar^2 k^2}{2 m_i} + \hbar c k} +
\mathcal{O}\left(\frac{\hbar\omega_0}{Mc^2} \right) \right]
\end{equation}
with $i=1,2$.} Here we neglect in the denominator the Doppler terms $\mathbf{Q}_0 \cdot \hbar
\mathbf{k}$ and $\mathbf{p} \cdot \hbar \mathbf{k}$, generated by
the identity $e^{i\textbf{k}\textbf{r}} H_{ho}(\textbf{p})
e^{-i\textbf{k}\textbf{r}} =  H_{ho}(\textbf{p}-\hbar \textbf{k})$.
They provide a finite correction of order $\frac{\hbar
\omega_0}{Mc^2} \sim 10^{-8}$. The leading contribution diverges
logarithmically. In the Bethe theory for the Lamb shift \cite{bethe}
exactly the same kind of divergency was encountered. The two
diverging mass-like terms $\delta m_i= \frac{4 \alpha}{3\pi} \hbar^2
\int kdk ( \frac{ \hbar^2 k^2}{2 m_i} + \hbar c k)^{-1}$ - here with
recoil effects included - stem  from the QED coupling of the free
particles 1 and 2 with the quantum vacuum and are therefore
naturally interpreted to be part of their intrinsic, observable
masses \cite{milonni}. {Two other diverging contributions are generated by the term $\alpha(0) \mathbf{E}_0 \times \mathbf{B}_0$ contained in $\mathbf{Q}_0$, see (\ref{class}), in the expression for $\Omega$ and add up to $ {-\delta m_1}/{M}- {\delta m_2}/{M}= {-\delta M}/{M}$. Adding up all diverging terms we obtain 
$$\mu^{-1}({\delta m_1}/{m_1} + {\delta
m_2}/{m_2} -  {\delta M}/{M}) = \mu^{-2}{\delta \mu} = - \delta
(1/\mu)$$} 
Since the static polarizability is proportional to
$1/\mu$, these UV-divergent corrections all disappear into the factor
$\alpha(0)$ of Eq.~(\ref{class}), which thus becomes defined in
terms of the observed reduced mass $\mu^*$ :
{\begin{equation} \nonumber
\mathbf{Q}_0 -  \delta (\frac{1}{\mu}) \frac{e^2}{\omega_0^2}\mathbf{E}_0 \times \mathbf{B}_0 = M\mathbf{v} - \frac{e^2}{\mu^* \omega_0^2}\mathbf{E}_0 \times \mathbf{B}_0 
\end{equation}}

All other terms generated by Eq.~(\ref{Kperturb}) are finite. In particular, the  term involving $\mathbf{p}_0$ in $\Omega$ generates the following  contribution to $\mathbf{K}$:
{\begin{align}\label{dvp01} \nonumber
  \mathbf{K }_1 &=  \alpha(0) \mathbf{E}_0 \times \mathbf{B}_0 \frac{ m_2-m_1}{2M} \frac{4 \alpha}{3\pi}  \\
&\times \lim_{\delta \downarrow 0}\int_{\delta}^\infty dk   \left( \frac{k}{  k^2/{2 } +  c k m_2/\hbar} - \frac{k}{  k^2/{2 } + \hbar c k m_1/\hbar } 
  \right) \nonumber \\
  &= -\alpha(0) \mathbf{E}_0 \times \mathbf{B}_0  \frac{4 \alpha}{3\pi} \frac{m_1-m_2 }{M}\log \frac{m_1}{m_2}.
\end{align}}
The UV divergency cancels out, and the integral is finite. Of course part of the k-integral enters the relativistic regime $\hbar k > m_ic$ in which the present theory is not valid. However, by subtracting and adding $2/k$ to both integrands in Eq.~(\ref{dvp01}) reveals two terms whose range of integration is typically $m_ic/\hbar$. This wavenumber was used by Bethe to cut off  his nonrelativistic theory for the Lamb shift \cite{bethe}.  We will thus adopt
 the final result in Eq.~(\ref{dvp01}) as a ``reasonable'' nonrelativistic estimate for the Casimir momentum. It is nevertheless  clear that a relativistic theory is required to get the complete picture
 for Casimir momentum.

All other cross-terms in Eq.~(\ref{Kperturb}) contain oscillating
exponential factors and converge rapidly for $k > 1/a$, i.e. stay in
the nonrelativistic regime . They generate a Casimir momentum that
is typically a factor $\sqrt{{\hbar \omega_0}/{\mu c^2}}\sim \alpha$
smaller. We obtain,
\begin{align}\label{cv}
\nonumber \mathbf{K}_2 &= \alpha(0) \mathbf{E}_0 \times \mathbf{B}_0  \,  \alpha \sqrt{\frac{\hbar \omega_0}{\mu c^2}} \\
& \times \left( - \frac{14 }{15 \sqrt{\pi}} +  \frac{2 }{3 \sqrt{\pi}} \left(\frac{\Delta m}{M}\right)^2  + \frac{8 }{3 \sqrt{\pi}}\frac{\mu}{M} \right)
\end{align}
When both masses are equal, $\mathbf{K}_2$ becomes the sole
contribution to Casimir momentum, {with a relative correction of order $\alpha^2$}. For $m_1 \gg m_2$ , $\mathbf{K}_1$
dominates. Since it is independent on details of the force between
the two particles, it is tempting to apply Eq.~(\ref{dvp01}) to the
hydrogen atom. With $\hbar \omega_0 = 10$ eV, $m_1=m_p$ and
$m_2=m_e$, $E_0=10^5 $ V/m and $B_0=17$ T, we find for the velocity
associated with the classical contribution~(\ref{class}) $v_{cl}
\approx 5 $ $\mu$m/s, and a QED correction  of 2 $\%$ in the same
direction; $\mathbf{K}_2$ yields a negligible correction of $0.01\,
\%$.

In conclusion, we have presented a non-relativistic quantum
electrodynamic theory for the total of an harmonic oscillator,
subject to external classical fields, and coupled to the
electromagnetic quantum vacuum. The most important conclusions of
this work are that Casimir momentum exists and that its UV
divergences are renormalizable. The theory shows it to be basically
a non-relativistic quantity, but that relativistic corrections are
likely to be significant, much like in the Lamb shift problem. To
our knowledge nor the classical contribution to magneto-electric
momentum, neither the QED correction have ever been observed.
\acknowledgments
We are indebted to Geert Rikken for many helpful discussions. This
work was supported by the contract PHOTONIMPULS ANR-09-BLAN-0088-01.

\end{document}